\documentclass[letterpaper]{article} 
\usepackage{aaai23}  
\usepackage{times}  
\usepackage{helvet}  
\usepackage{courier}  
\usepackage[hyphens]{url}  
\usepackage{graphicx} 
\usepackage{natbib}  
\usepackage{caption} 
\usepackage{algorithm}
\usepackage{algorithmic}

\usepackage{newfloat}
\usepackage{listings}
\usepackage{subfig}
\DeclareCaptionStyle{ruled}{labelfont=normalfont,labelsep=colon,strut=off} 
\floatstyle{ruled}
\newfloat{listing}{tb}{lst}{}
\floatname{listing}{Listing}
\title{Just Another Day on Twitter: A Complete 24 Hours of Twitter Data}

\author{
    Jürgen Pfeffer\textsuperscript{\rm 1}, 
    Daniel Matter\textsuperscript{\rm 1},
    Kokil Jaidka\textsuperscript{\rm 2},
    Onur Varol\textsuperscript{\rm 3},
    Afra Mashhadi\textsuperscript{\rm 4},
    Jana Lasser\textsuperscript{\rm 5, \rm 15},
    Dennis Assenmacher\textsuperscript{\rm 6}, 
    Siqi Wu\textsuperscript{\rm 7},
    Diyi Yang\textsuperscript{\rm 8},
    Cornelia Brantner\textsuperscript{\rm 9},
    Daniel M. Romero\textsuperscript{\rm 7},
    Jahna Otterbacher\textsuperscript{\rm 10},
    Carsten Schwemmer\textsuperscript{\rm 11},
    Kenneth Joseph\textsuperscript{\rm 12},
    David Garcia\textsuperscript{\rm 13}, 
    Fred Morstatter\textsuperscript{\rm 14}
}
\affiliations{
    \textsuperscript{\rm 1}School of Social Sciences and Technology, Technical University of Munich, 
    \textsuperscript{\rm 2}Centre for Trusted Internet and Community, National University of Singapore,
    \textsuperscript{\rm 3}Sabanci University,
    \textsuperscript{\rm 4}University of Washington (Bothell),
    \textsuperscript{\rm 5}Graz University of Technology,
    \textsuperscript{\rm 6}GESIS--Leibniz Institute for the Social Sciences,
    \textsuperscript{\rm 7}University of Michigan,
    \textsuperscript{\rm 8}Stanford University,
    \textsuperscript{\rm 9}Karlstad University,
    \textsuperscript{\rm 10}Open University of Cyprus \& CYENS CoE,
    \textsuperscript{\rm 11}Ludwig Maximilian University of Munich,
    \textsuperscript{\rm 12}University at Buffalo,
    \textsuperscript{\rm 13}University of Konstanz,
    \textsuperscript{\rm 14}Information Sciences Institute, University of Southern California
    \textsuperscript{\rm 15}Complexity Science Hub Vienna
}

\usepackage{bibentry}

\begin{document}

\maketitle

\begin{abstract}
At the end of October 2022, Elon Musk concluded his acquisition of Twitter. In the weeks and months before that, several questions were publicly discussed that were not only of interest to the platform's future buyers, but also of high relevance to the Computational Social Science research community. For example, how many active users does the platform have? What percentage of accounts on the site are bots? And, what are the dominating topics and sub-topical spheres on the platform? In a globally coordinated effort of $80$ scholars to shed light on these questions, and to offer a dataset that will equip other researchers to do the same, we have collected \emph{all} 375 million tweets published within a $24$-hour time period  starting on September 21, 2022. To the best of our knowledge, this is the first complete 24-hour Twitter dataset that is available for the research community. With it, the present work aims to accomplish two goals. First, we seek to answer the aforementioned questions and provide descriptive metrics about Twitter that can serve as references for other researchers. Second, we create a baseline dataset for future research that can be used to study the potential impact of the platform's ownership change.
\end{abstract}

\section{Introduction}
On March 21, 2006, Twitter's first CEO Jack Dorsey sent the first message on the platform. In the subsequent 16 years, close to 3 trillion tweets have been sent.\footnote{While we do not have an official source for this number, it represents an educated guess from a collaboration of dozens of scholars of Twitter.}
Roughly two-thirds of these have been either removed from the platform because the senders deleted them or because the accounts (and all their tweets) have been banned from the platform, have been made private by the users, or are otherwise inaccessible via the historic search with the v2 API endpoints. By utilizing Twitter's \emph{count/all API} and the approaches described in this article, we estimate that about 900 billion public tweets were on the platform when Elon Musk acquired Twitter in October 2022 for \$44B \footnote{https://www.nytimes.com/2022/10/27/technology/elon-musk-twitter-deal-complete.html}.

Besides its possible economic value, Twitter has been instrumental in studying human behavior with social media data and the entire field of Computational Social Science (CSS) has heavily relied on data from Twitter. At the AAAI International Conference on Web and Social Media (ICWSM), in the past two years alone (2021-2022), over 30 scientific papers analyzed a subset of Twitter for a wide range of topics ranging from public and mental health analyses to politics and partisanship. Indeed, since its emergence, Twitter has been described as a digital \textit{socioscope} (i.e., social telescope) by  researchers in fields of social science~\cite{mejova2015twitter}, ``a massive antenna for social science that makes visible both the very large (e.g., global patterns of communications) and the very small (e.g., hourly changes in emotions)''.
Beyond CSS, there is increasing use of Twitter data for training large pre-trained language models in the field of natural language processing and machine learning, such as Bernice \cite{deluciabernice}, where 2.5 billion tweets are used to develop representations for Twitter-specific languages, and TwHIN-BERT \cite{zhang2022twhin} that leverages 7 billion tweets covering over 100 distinct languages to model short, noisy, and user-generated text. 

Although Twitter data has fostered interdisciplinary research across many fields and has become a ``model organism'' of big data, scholarship using Twitter data  has also been criticized for various forms of bias that can emerge during analyses~\cite{tufekci2014big}. One major challenge giving rise to these biases is getting access to data and knowing about data quality and possible data biases~\cite{Ruths2014,Gonzalez2014,olteanu2019social}. While Twitter has long served as one of the most collaborative big social media platforms in the context of data-sharing with academic researchers, there nonetheless exists a lack  of  transparency in sampling procedures and possible biases created from technical artifacts~\cite{Morstatter2013, Pfeffer2018Tampering}. These unknown biases may  jeopardize research quality. At the same time, 
access to unfiltered/unsampled Twitter data is nearly impossible to access, and thus the above-mentioned studies, as well as thousands of others, still retain unknown and potentially significant biases in their use of sampled data. Only a few studies have tried to collect complete Twitter data samples. For instance, Geenen et al. \citeyearpar{Geenen2016} have utilized a third-party data provider to collect all Dutch-speaking tweets over a period of one week. In 2010, Kwak et al. \citeyearpar{Kwak2010} have crawled the entire Twitter site to get 41.7 million user profiles and 1.47 billion following connections. At around the same time, Cha et al. \citeyearpar{Cha2010} found $\sim$55 million accounts and 1.96 billion following connections.

\paragraph{Contributions.} The data collection efforts presented in this paper were driven by a desire to address these concerns about sampling bias that exist because of the lack of a complete sample of Twitter data. Consequently, the main contribution of this article is to create the first \emph{complete} dataset of 24 hours on Twitter and make these tweets available via future collaborations with the authors and contributors of this article. The dataset collected and described here can be used by the research community to:
\begin{itemize}
    \item Promote a better understanding of the communication dynamics on the platform. For example, it can be used to answer questions like, how many active (posting) accounts are on the platform? And, what are the dominating languages and topics?
    \item Create a set of descriptive metrics that can serve as references for the research community and provide context to past and present research papers on Twitter. 
    \item Provide a baseline for the situation \emph{before} the recent sale of Twitter. With the new ownership of Twitter, platform policies as well as the company structures are under significant change, which will create questions about whether previous Twitter studies will be still valuable references for future studies. 
\end{itemize}

In the following sections, we describe the data collection process and provide some descriptive analyses of the dataset. We also discuss ethical considerations and data availability. 

\section{Data}
\paragraph{Data Collection.}
We have collected $24$ hours of Twitter data from September 20, 15:00:00 UTC to September 21 14:59:59 UTC. The data collection was accomplished by utilizing the Academic API \cite{Pfeffer2023Academic} that is free and openly available for researchers. The technical setup of the data collection pipeline was dominated by two major challenges: First, how can we avoid---at least to a satisfying extent---a temporal bias in data collection? Second, how can we get a \emph{good} representation of Twitter? In the following, these two aspects are discussed in more detail.

\paragraph{What is a complete dataset?}
What does \emph{complete} mean when we want to collect a day's worth of Twitter data? It has been shown previously that the availability of tweets fluctuates, especially in the first couple of minutes \cite{Pfeffer2023Academic}---people might delete their tweets because of typos, tweets might be removed because of violations of terms of service, etc. To reduce this initial uncertainty, we have decided to collect the data 10 minutes after the tweets were sent. Consequently, this dataset does not include all tweets that were sent on the collection day but instead tries to create a somewhat \emph{stable} representation of Twitter.

\paragraph{Avoiding temporal collection bias.}
We wanted to collect a set of tweets close to the time when they were created. However, collecting data takes time, which can introduce possible temporal bias, e.g., if we want to collect data from the previous hour and the data collection job takes three hours, then the data that is collected at the end of the collection job will be much older (with potentially more tweet removals) than the data that is collected at the beginning. To tackle this challenge, we have split the day into 86,400 collection tasks, each consisting of 1 second of Twitter activity. The collection of every second of data started exactly 10 minutes after the data creation time. Because the data collection of a second took more than a minute during peak times, we have distributed the workload to 80 collection processes, i.e., Academic API tokens, in order to avoid backlogs.

\paragraph{Data collection queries.}
The backbone of our data collection effort is a query that---to the best of our knowledge---is able to collect \emph{ALL} tweets within a specified time frame. The query is based on the following three aspects. First, the Academic API allows for negative selectors to limit a search query, e.g., ``indictment -trump'' will return a tweet including the term ``indictment'' only if the tweet does not include the term ``trump''. While simple ``A and not A'' queries are not allowed to collect all possible Tweets, a negative selection is possible when combined with a language selector. The query ``-trump lang:de'' will return all German tweets that do not include the term ``trump''. If we now replace the term ``trump'' with a long random string, we will receive \emph{all} German tweets. Second, Twitter assigns a language code to \emph{every} tweet\footnote{We have verified this with 1\% Sample API data.} (including ``und'' for undefined). Consequently, we can construct a long \emph{OR}-condition with all possible languages: (lang:am OR lang:ar OR ...). We can get the list of all currently used Twitter languages by requesting a non-existing language, e.g., ``-trump lang:test'' will return an error including a list of all possible languages---currently, there are 74 possible languages. Combined with the exclusion of the long random string, this will query all tweets on the platform. Third, it is possible to limit the time frame for data collection to a single second, e.g., by setting the start to ``2022-09-21T12:00:00Z'' and the end to ``2022-09-21T12:00:01Z''.

\begin{figure}[t]
\centering
\includegraphics[width=0.95\linewidth]{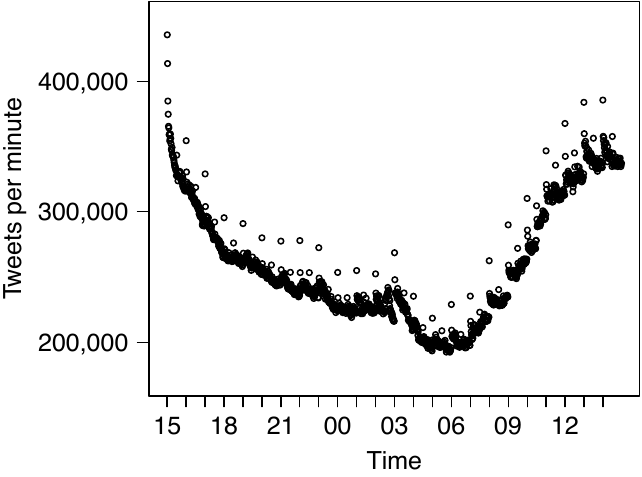}
    \caption{Tweets per minute over the 24-hour collection period, time in UTC.}
    \label{fig:count}
\end{figure}

\paragraph{Number of tweets.}
With the above-described process, we have collected 374,937,971 tweets within the 24-hour time span. On average, this amounts to 4,340 [2,989 -- 8,955] tweets per second. Fig. \ref{fig:count} plots the number of tweets per Minute (avg=260,374, min=192,322, max=435,721). The data collection started at 15:00 UTC, when almost the entire Twitter world is awake. Then, we can see from Japan to Europe time zone after time zone getting off the platform. While Europe and the Americas are sleeping, Asia keeps the number of tweets at around 200,000. Starting at 7:00 UTC, Europe is getting active again, followed by the Americas from East to West. Another astonishing observation of this time series is that the first minute of every hour has on average 15.5\% more tweets than the minute before---most likely due to bot activities and other timed tweet releases, e.g., news.

\section{Descriptive Analyses}
\subsection{Active Users}
The 375 million tweets in our dataset were sent by 40,199,195 accounts. While the publicly communicated numbers of users of a platform are often based on the number of active and passive visitors, we can state that Twitter has (or at least had on our observed day) 40 million active contributors who have sent at least one tweet. Less than 100 accounts have created about 1\% (=3.5M) tweets. ${\sim}175,000$ accounts (0.44\%) created 50\% of all tweets.

\begin{table}[b!]
\centering
\footnotesize
\caption{Distribution of user activity}
\begin{tabular}{rrr}
\hline
\textbf{\% Total Tweets} & \textbf{\% Total Users} & \textbf{Min. no. of Tweets}\\
\hline
1\% &     0.00023\% &   2,267\\
10\% &    0.01199\%  &  465\\
25\%  &   0.07284\%   & 152\\
50\%   &  0.43526\%    & 39\\
75\%    & 1.70955\%    & 11\\
90\%     &4.18836\%    & 3\\
\hline
\end{tabular}
\label{tab:datasetstats} 
\end{table}

These numbers are not surprising when we consider that $>95\%$ of active accounts have sent one or two tweets. However, these numbers lend more nuance to recent reports from the Pew Research Center, which reported that while the majority of Americans use social media, approximately 97\% of all tweets were posted by 25\% of the users~\citep{socialmedia2021pew}. In fact, our dataset suggests that worldwide, the numbers may be more skewed than previously suggested.

\subsection{User metrics}

\begin{figure}[t]
    \centering
    \includegraphics[width=0.97\linewidth]{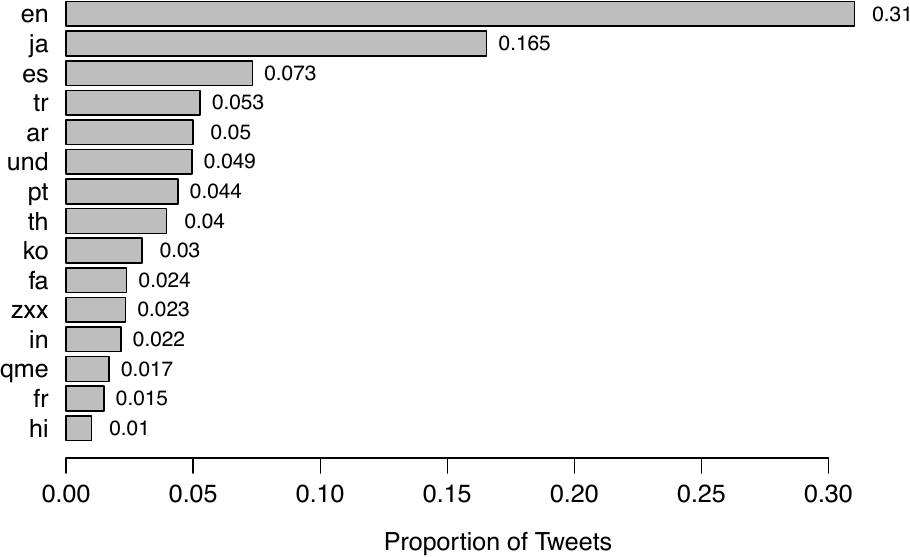}
    \caption{All languages occurring in at least 1\% of the tweets.}
    \label{fig:lang}
\end{figure}

\paragraph{Followers.}
The active accounts on our day of Twitter data have a mean of 2,123 followers (median=99).  We can find six accounts with more than 100 million followers (max=133,301,854), and 427/8,635 accounts with more than 10/1 million followers. Exactly 50\% of accounts that were active on our collection day have less than 100 followers.

\paragraph{Following.}
These accounts follow much fewer other accounts: mean=547, median=197, range: 0--4,103,801. Interestingly, there are 2,377 accounts that follow more than 100,000 other accounts. One-third of accounts follow less than 100 accounts, but only 1.7\% of accounts follow zero other accounts.

\paragraph{Listed.}
Lists are a Twitter feature for users to organize accounts around topics and filter tweets. While there is little evidence that lists are used widely on the platform, this feature might be useful for getting an impression about the number of \emph{interesting} content creators on the platform. The 40 million active accounts in our dataset are listed (i.e., number of lists that include a user) in 0 to 3,086,443 lists (mean=10.1, median=0). 1,692/46,139 accounts are in lists of at least 10,000/1,000 accounts.

\paragraph{Tweets sent.}
The user information of the tweet metadata also includes the number of tweets that a user has sent---or at least how many of those tweets are still available on Twitter. The sum of the sent tweets variable of all 40 million accounts is ${\sim}$404 billion (mean=9,704, median=1,522). If we assume that our initial estimate of having 900 billion tweets on the platform at the time of data collection is somewhat correct, the accounts active in our dataset have contributed ${\sim}$45\% of all of the available tweets over the entire lifetime of Twitter.

\paragraph{Verified accounts.} At the time of our data collection, we can identify 221,246 verified accounts among the 40 million active users.

\subsection{Tweets and retweets}
79.2\% of all tweets refer to other tweets, i.e. they are retweets or quotes of or replies to other tweets. Consequently, 20.8\% of the tweets in our dataset are original tweets. The tweets with references are of the following types: 50.7\% retweets, 4.3\% quotes, 24.2\% replies, i.e. half of all tweets are retweets and a fourth are replies.

\paragraph{Retweeted and liked.}
Studying the retweet and like numbers from the tweets' metadata has created little insight since the top retweeted tweets are very old tweets that have been retweeted by chance on our collection day. Furthermore, we can see the number of likes only for tweets that have been tweeted and retweeted. In any case, the retweeted number is interesting---the 374 million tweets have been retweeted 401 billion times. In other words, significant parts of historic Twitter get retweeted on a daily basis.

\subsection{Languages}
Twitter annotates a language variable for every tweet. Fig. \ref{fig:lang} shows those languages that were annotated on at least 1\% of our dataset. Together, these 15 languages make up 92.5\% of all tweets. Besides the most common languages on Twitter, we can also find interesting language codes in this list: \emph{und} stands for undefined and represents tweets for which Twitter was not able to identify a language; \emph{qme} and \emph{zxx} seem to be used by Twitter for tweets consisting of only media or a Twitter card. 

\subsection{Media} 
There are 112,779,266 media attachments in our data collection (76.9\% photos, 20.7\% videos, 2.4\% animated GIFs), of which 37,803,473 have unique media keys (83.8\% photos, 10.0\% videos, 6.2\% animated GIFs).

\subsection{Geo-tags} 
\begin{figure}
    \centering
    \includegraphics[width=\linewidth]{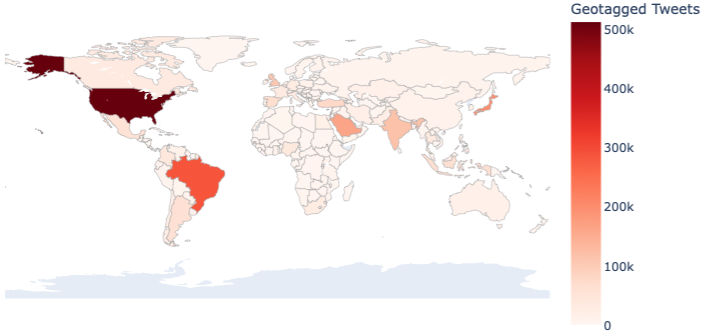}
    \caption{Choropleth map of the geo-tagged tweets across the world. }
    \label{fig:geo}
\end{figure}

We found only  0.5\% of tweets to be geo-tagged. This is not surprising as previous works have shown that the percentage of geo-tagging in Twitter has been declining~\cite{ajao2015survey}. Fig.~\ref{fig:geo} shows the distribution of the geo-tagged tweets across the world, with USA (20\%), Brazil (11\%), Japan (8\%), Saudi Arabia (6\%) and India (4\%) being the top five countries. 

\subsection{Estimating prevalence of bot accounts}

Twitter has a pivotal role in public discourse and entities that are after power and influence often utilize this platform through social bots and other means of automated activities. Since the early days of Twitter, researchers have been studying bot behavior, and it has become an active research area~\cite{ferrara2016rise,cresci2020decade}. The first estimation of bot prevalence on Twitter indicates that 9-15\% of Twitter accounts exhibit automated behavior~\cite{varol2017online}, while others have observed significantly higher percentages of tweets produced by bot-likely accounts on specific discourses~\cite{uyheng2021computational,antenore2022comparative}. One major challenge in estimating bot prevalence is the variety of definitions, datasets, and models used for detection~\cite{varol2022should}.

\begin{figure*}[th!]%
    \centering
    \subfloat[\centering]{{
        \includegraphics[width=0.33\textwidth]{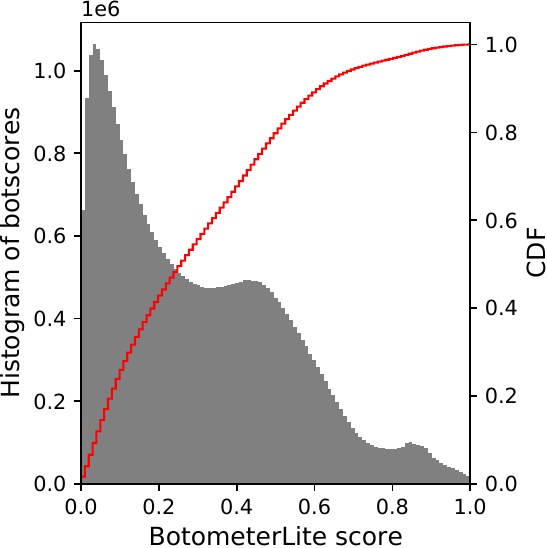} 
        \label{fig:botscores} }}%
    \subfloat[\centering]{{
        \includegraphics[width=0.32\textwidth]{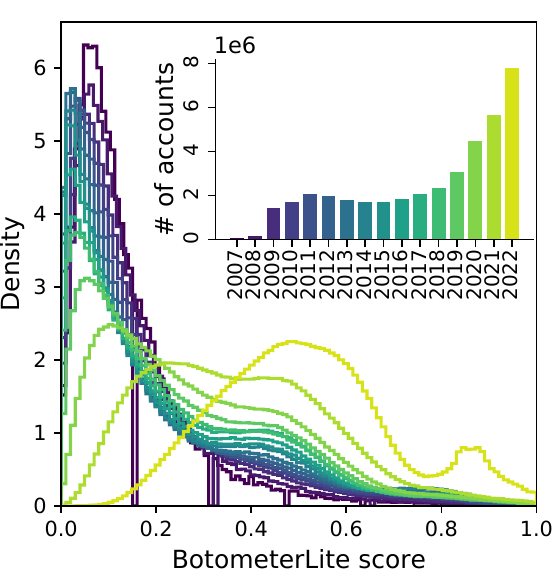} 
        \label{fig:botscores-age}
    }}%
    \subfloat[\centering]{{
        \includegraphics[width=0.32\textwidth]{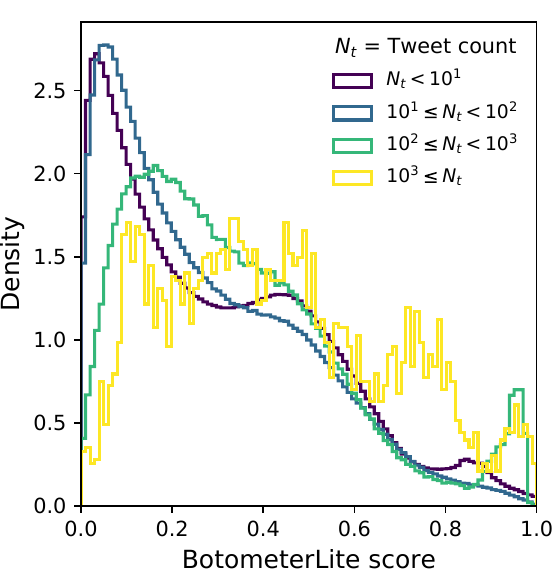} 
        \label{fig:botscores-tweets}
    }}%
    \caption{BotometerLite scores distribution: (a) histogram and cumulative distribution, (b) by account age, (c) by tweet counts in our dataset.}%
    \label{fig:example}%
\end{figure*}

In this study, we employed BotometerLite~\cite{yang2020scalable}, a scalable and light-weight version of the Botometer~\cite{sayyadiharikandeh2020detection}, for computing bot scores for unique accounts in our collection. In Fig. \ref{fig:botscores}, we present the distribution of bot scores and nearly 20\% of the 40 million active accounts have scores above 0.5 suggesting bot-likely behavior.

While identification of bots is a complex and possibly controversial challenge, plotting the distributions of BotometerLite scores grouped by account age in Fig. \ref{fig:botscores-age} suggests the proportions of accounts that show bot-like behavior has increased dramatically in recent years. This result may also suggest that the longevity of simpler bot accounts is limited and they are no longer active on the platform. In Fig. \ref{fig:botscores-tweets}, we also present the distribution of bot scores for different rates of activities in our dataset. Accounts that have over 1,000 posts exhibit higher rates of bot-like behaviors. 

It is important to mention that accounts studied in this paper were identified due to their content creation activities. Our collection cannot capture passive accounts that are simply used to boost follower counts without visible activity on tweet streams. Fair assessment of bot prevalence is only possible with complete access to Twitter's internal database; since activity streams, network data, and historical tweet archives can capture different sets of accounts \cite{varol2022should}.

\subsection{Content on Twitter}
The top 500 hashtags occurred 81,468,508 times in the tweets. Via manual inspection, we were able to identify the meaning of 95\% of these top hashtags. They can be aggregated into ten the categories.

Table~\ref{tab:hashtags} suggests that a large proportion of tweets referred to entertainment, which together comprised about 30\% of tweets. These included mentions of celebrities (25.5\%) and other entertainment-related tweets (5.4\%) such as mentions of South Korean boy band members, and other references to music, movies, and TV shows. Our data collection time window occurred during Fall/Winter 2022, when the world was discussing the protests in Iran after the death of Mahsa Amini. Therefore, the Iranian protests also comprised a large proportion of the hashtag volume at 16.6\%.


\begin{table}[b]
\caption{The categories of the top 500 hashtags in the dataset}

\begin{tabular}{lrrr}

\hline
\textbf{Category} & \multicolumn{1}{l}{\textbf{Hashtags}} & \multicolumn{1}{l}{\textbf{Occurrence}} & \multicolumn{1}{l}{} \\ \hline
Celebrities      & 159 & 20,809,742 & 25.5\%  \\
Sex              & 104 & 20,529,196 & 25.2\%  \\
Iranian Protests & 15  & 13,488,295 & 16.6\%  \\
Entertainment    & 45  & 4,392,227  & 5.4\%   \\
Advertisement    & 32  & 4,644,540  & 5.7\%   \\
Politics         & 38  & 3,858,550  & 4.7\%   \\
Finance          & 30  & 3,549,107  & 4.4\%   \\
Games            & 21  & 3,348,128  & 4.1\%   \\
Other            & 31  & 2,672,291  & 3.3\%   \\
Unknown          & 25  & 4,176,432  & 5.1\%   \\ \hline
Sum              & 500 & 81,468,508 & 100.0\% \\ \hline
\end{tabular}
\label{tab:hashtags}
\end{table}


\begin{figure*}[t!]
    \centering
    \includegraphics[width=0.96\textwidth]{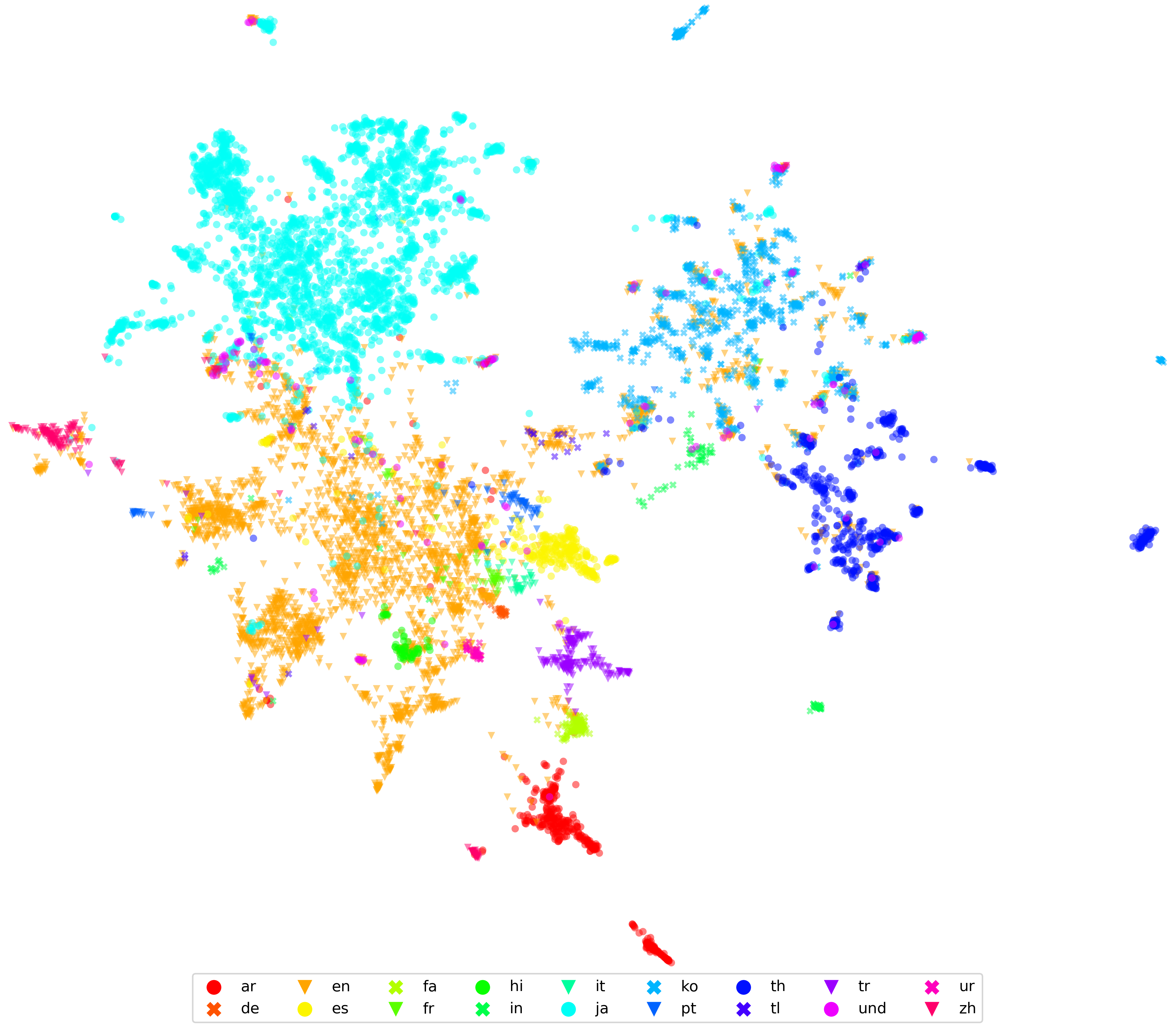}
    \caption{MDS of top 10,000 hashtags based on co-usage by same accounts; colors represent dominant language in tweets using a hashtag.}
    \label{fig:topics}
\end{figure*}

Finally, and perhaps surprisingly, the category \emph{sex} comprised over a quarter of all content covered by the top hashtags, and was almost completely related to escorts. ``Other" topics reflect that on ``regular" Twitter days, sports, tech, and art may take up only about 3.3\% of Twitter volume.


Fig. \ref{fig:topics} is a hashtag visualization that attempts to provide an overview of the entire content on Twitter. We first removed all tweets from accounts with more than 240 tweets (=10\% of the max. daily allowed number of tweets) to reduce the noise from bots using random trending hashtags. From the remaining tweets, we extracted the 10,000 most often used hashtags in our dataset and created a hashtag similarity matrix with the number of accounts that have used a pair of two hashtags on the day of data collection. Every element in Fig. \ref{fig:topics} represents a hashtag. The position is the result of Multidimensional Scaling (MDS) and the color shows the dominant language that was used in the tweets with the particular hashtag. In this figure, we can see how languages separate the Twitter universe but that there are also topical sub-communities within languages.

 \section{Discussion and Potential Applications}
Twitter is a social media platform with a worldwide userbase. Open access to its data also makes it attractive to a large community of researchers, journalists, technologists, and policymakers who are interested in examining social and civic behavior online. Early studies of Twitter explored who says what to whom on Twitter~\cite{wu2011says}, characterizing its primary use as a communication tool. Other early work mapped follower communities through ego networks~\cite{gruzd2011imagining}. However, Twitter has since expanded into its own universe, with a plethora of users, uses, modalities, communities, and real-life implications. Twitter is increasingly the source of breaking news, and many studies from the U.S. and Europe have reported that Twitter is one of the primary sources of news for their citizens. Twitter has been used for political engagement and citizen activism worldwide. During the COVID-19 pandemic, Twitter even assumed the role of the official mouthpiece and crisis communication tool for many governments to contact their citizens, and from which citizens could seek help and information.

Fig.~\ref{fig:geo} confirms prior reports that geotagging practices are limited in many low- and middle-income countries~\cite{malik2015population}; however, this should not deter scholars from exploring alternative methods of triangulating the location of users~\cite{schwartz2013characterizing}, and creating post-stratified estimates of regional language use~\cite{jaidka2020estimating,giorgi2022correcting}. In prior studies, the difficulties in widespread data collection and analyses have so far implied that most answers are based on smaller samples (usually constrained by geography, for convenience) of a burgeoning Twitter population. Fig. \ref{fig:topics} and Table \ref{tab:hashtags} also impressively illustrate that Twitter is about so much more than US politics.

We hope that our dataset is the first step in creating alternatives for conducting a representative and truly inclusive analysis of the Twitterverse. Temporal snapshots are invaluable to map the national and international migration patterns that increasingly blur geopolitical boundaries~\cite{zagheni2014inferring}.

The increasing popularity of Twitter has led it into issues of scale, where its moderation can no longer check the large proportion of bots on the platform. Our findings in Fig.~\ref{fig:example} indicate that the infestation of bots may be more pernicious than previously imagined. We are especially concerned that the escalation of the war on Ukraine by Russia may reflect a spike (in our dataset) in the online activity of bots from Russia operated either by the Russian government or its allied intelligence agencies~\cite{badawy2018analyzing}. These and other bots serve to amplify trending topics and facilitate the spread of misinformation (though, perhaps, at a rate less than humans do~\cite{vosoughi2018spread}). They may also misuse hashtags to divert attention away from social or political topics~\cite{earl2022digital,broniatowski2018weaponized} or strategically target influential users \cite{shao2018spread,varol2020journalists}. We hope that our work will spur more studies on these topics, and we welcome researchers to explore our data. 

By observing bursts of discussions around politically charged events and characterizing the temporal spikes in Twitter topics, we can better rationalize how our experience of Twitter as a political hotbed differs from the simplified understanding of the American Twitter landscape reported in~\citet{mukerjee2022political}, which suggested that politics is largely a sideshow on Twitter. It is worth considering that these politically active users may not be representative of social media users at large~\cite{socialmedia2021pew,wojcieszak2022most}.

Twitter is also under scrutiny for how its platform governance may conflict with users' interests and rights~\cite{van2018platform}. Concerns have been raised about alleged biases in the algorithmic amplification (and deamplification) of content, with evidence from France, Germany, Turkey, and the United States, among other countries~\cite{majo2021role,tanash2015known,jaidka2023silenced}. Other scholars have also criticized Twitter's use as a censorship weapon by governments and political propagandists worldwide~\cite{varol2016spatiotemporal,elmas2021dataset,jakesch2021trend}. They, and others, may be interested in examining the trends in the enforcement of content moderation policies by Twitter. 

Besides answering questions of data, representativeness, access, and censorship, we anticipate that our dataset is suited to explore the temporal dynamics of online (mis)information in the following directions:
\begin{itemize}
    \item \textbf{Content characteristics}: We have provided a high-level exploration of the topics on Twitter. However, more can be done with regard to understanding users' concerns and priorities. While hashtags act as signposts for the broader Twitter community to find and engage in topics of mutual interest~\cite{cunha2011analyzing}, tweets without hashtags may offer a different understanding of Twitter discourse, where users may engage in more interpersonal discussions of news, politics, and sports than the numbers suggest~\cite{rajadesingan2021political}. 
    \item \textbf{Patterns of information dissemination:} Informational exchanges occurring on Twitter can overcome spatio-temporal limitations as they essentially reconfigure user connections to create newly emergent communities. However, these communities may vanish as quickly as they are created, as the lifecycle of a tweet determines how long it continues to circulate on Twitter timelines. To the best of our knowledge, no prior research has reported on the average ``age" of a tweet, and we hope that a 24-hour snapshot will enable us to answer this question empirically. 
    \item \textbf{Content moderation and fake news}: Prior research suggests that 0.1\% of Twitter users accounted for 80\% of all fake news sources shared in the lead-up to a US election~\cite{grinberg2019fake}. However, we expect there to be cross-lingual differences in this distribution, especially for low- or under-resourced languages with fewer open tools for fact-checking. Similarly, we expect that the quality of moderation and hate speech will vary by geography and language, and recommend the use of multilingual large language models to explore these trends (with attention to persisting representativeness caveats~\cite{wu2020all}).  
    \item \textbf{Mass mobilization}: Twitter is increasingly the hotbed of protest, which has led to some activists donning the role of ``movement spilloverers"~\cite{zhou2021longitudinal} or serial activists~\cite{bastos2016serial} who broker information across different online movements, thereby acting as key coordinators, itinerants, or gatekeepers in the exchange of information. Such users, as well as the constant communities in which they presumably reside~\cite{chowdhury2022constant}, may be easier to study through temporal snapshots, as facilitated by this dataset.
    \item \textbf{Echo chambers and filter bubbles}: On Twitter, algorithms can affect the information diets of users in over 200 countries, with an estimated 396.5 million monthly users~\cite{Statista2022}. Recent surveys of the literature have considered the evidence on how platforms' designs and affordances influence users behaviors, attitudes, and beliefs~\cite{gonzalez2022social}. Studies of the structural and informational networks based on snapshots of Twitter can offer clues to solving these puzzles without the constraints of data selection.
    
\end{itemize}

\section{Ethics Statement and Data Availability}
 
 \paragraph{Ethics statement.}
We acknowledge that privacy and ethical concerns are associated with collecting and using social media data for research. However, we took several steps to avoid risks to human subjects since participants no longer opt into being part of our study, in a traditional sense~\cite{zimmer2020but}. In our analysis, we only studied and reported population level, and aggregated observations of our dataset. We share publicly only the tweet IDs with the research community to account for privacy issues and Twitter's TOS. For this purpose, we use a data sharing and long-term archiving service provided by GESIS - Leibniz Institute for the Social Sciences, a German infrastructure institute for the social sciences \footnote{\url{https://www.gesis.org/en/data-services/share-data}}.

With regards to data availability, this repository adheres to the FAIR principles~\cite{wilkinson2016fair} as follows:
\begin{itemize}
\item \textbf{Findability}: In compliance with Twitter's terms of service, only tweet IDs are made publicly available at DOI: \url{https://doi.org/10.7802/2516}. A unique Document Object Identifier (DOI) is associated with the dataset. Its metadata and licenses are also readily available.
\item \textbf{Accessibility}: The dataset can be downloaded using standard APIs and communications protocols (the REST API and OAI-PMH).   
\item \textbf{Interoperability}: The data is provided in raw text format. 
\item \textbf{{Reusability}}: The CC BY 4.0 license implies that researchers are free to use the data with proper attribution. 
\end{itemize}

In light of the recent changes to Twitter's APIs, we expect significant limitations when accessing tweets. Consequently, we want to invite the broader research community to approach one or more of the authors and collaborators (see Acknowledgments) of this paper with research ideas about what can be done with this dataset. We will be very happy to collaborate with you!

\section{Acknowledgments}
The data collection effort described in this paper could not have been possible without the great collaboration of a large number of scholars, here are some of them (in random order): Chris Schoenherr, Leonard Husmann, Diyi Liu, Benedict Witzenberger, Joan Rodriguez-Amat, Florian Angermeir, Stefanie Walter, Laura Mahrenbach, Isaac Bravo, Anahit Sargsyan, Luca Maria Aiello, Sophie Brandt, Wienke Strathern, Bilal Çakir, David Schoch, Yuliia Holubosh, Savvas Zannettou, Kyriaki Kalimeri.

\bibliography{aaai23}

\end{document}